\documentclass[preprint,aps,amsmath,nofootinbib,superscriptaddress]{revtex4}
\usepackage{graphicx}
\usepackage{bm}
\usepackage{epsfig}

\newif\ifpdf
\ifx\pdfoutput\undefined
\pdffalse 
\else
\pdfoutput=1 
\pdftrue
\fi

\def\Dslash{D\!\!\!\!\slash}
\def\cDslash{{\cal D}\!\!\!\!\slash}
\def\Dcslash{{\cal D}\!\!\!\!\slash}

\def\SppP{{\cal {P\!\!\!\!\hspace{0.04cm}\slash}}_\perp}

\def\nslash{n\!\!\!\slash}

\def\bnslash{\bar n\!\!\!\slash}

\def\epsslash{\varepsilon\!\!\!\slash}

\def\Delslash{\Delta\!\!\!\slash}

\def\Aslash{A\!\!\!\slash}
\def\OMIT#1{}

\newcommand{\mklg}[1]{\mbox{\large $#1$}}
\newcommand{\mklga}[1]{\mbox{\Large $#1$}}

\newcommand{\nn}{\nonumber} 

\newcommand{\bn}{{\bar n}}
\newcommand{\bea}{\begin{eqnarray}}
\newcommand{\eea}{\end{eqnarray}}

\newcommand{\bnP}{\bar {\cal P}}

\newcommand{\cP}{{\cal P}}

\newcommand{\mcdot}{\!\cdot\!}

\newcommand{\cD}{{\cal D}}

\begin{document}
\ifpdf
\DeclareGraphicsExtensions{.pdf, .jpg}
\else
\DeclareGraphicsExtensions{.eps, .jpg}
\fi


\preprint{ \vbox{\hbox{UCSD/PTH 02-07} \hbox{INT-PUB-02-36} \hbox{DUKE-TH-02-220}
}}

\title{\phantom{x}\vspace{0.5cm} 
Reparameterization Invariance for Collinear Operators
\vspace{0.5cm} }

\author{Aneesh V. Manohar}
\affiliation{Department of Physics, University of California at San Diego,
	La Jolla, CA 92093 \footnote{Electronic address: manohar@wigner.ucsd.edu,
	pirjol@bose.ucsd.edu}}

\author{Thomas Mehen}
\affiliation{Department of Physics, Duke University, Durham NC 27708 
\footnote{Electronic address: mehen@phy.duke.edu}}
\affiliation{Jefferson Laboratory 12000 Jefferson Ave. Newport News VA 23606}
	
\author{Dan Pirjol}
\affiliation{Department of Physics, University of California at San Diego,
	La Jolla, CA 92093 \footnote{Electronic address: manohar@wigner.ucsd.edu,
	pirjol@bose.ucsd.edu}}

\author{Iain W. Stewart\vspace{0.4cm}}
\affiliation{Department of Physics, University of California at San Diego,
	La Jolla, CA 92093 \footnote{Electronic address: manohar@wigner.ucsd.edu,
	pirjol@bose.ucsd.edu}}
\affiliation{Institute for Nuclear Theory,  University of Washington, Seattle, 
	WA 98195 \footnote{Electronic address: iain@phys.washington.edu}
	\vspace{0.5cm}}


\begin{abstract}
\vspace{0.5cm}
\setlength\baselineskip{18pt}

We discuss restrictions on operators in the soft-collinear effective theory
(SCET) which follow from the ambiguity in the decomposition of collinear momenta
and the freedom in the choice of light-like basis vectors $n$ and
$\bn$. Invariance of SCET under small changes in $n$ and/or $\bn$ implies a
symmetry of the effective theory that constrains the form of allowed operators
with collinear fields. The restrictions occur at a given order in the power
counting as well as between different orders. As an example, we present the
complete set of higher order operators that are related to the collinear quark
kinetic term.

\end{abstract}

\maketitle


\newpage

\setlength\baselineskip{15pt}

Strong interaction processes involving highly energetic particles can be
described within an effective field theory framework known as the Soft-Collinear
Effective Theory (SCET)~\cite{bfl,bfps,cbis,bpssoft}. SCET has been
used to simplify proofs of classical factorization theorems~\cite{bfprs},
provide the first all-orders proof of factorization in $\bar B \rightarrow D
\pi$ decays~\cite{bps}, and facilitate the resummation of Sudakov logarithms in
$B$ and $\Upsilon$ decays~\cite{bfl,bfps,xian}.

Suppose a set of light hadrons are created in a hard-scattering process.  The
hadrons are assumed to have a large energy $Q \gg \Lambda_{\rm QCD}$, and
invariant mass $\sim\Lambda_{\rm QCD}$. To a good approximation each light
hadron is composed of constituents nearly collinear to a light-like vector
$n$. To decompose collinear momenta it is necessary to define an auxiliary
light-like vector for an orthogonal direction, $\bn$, such that {$n\cdot\bn=2$}.
If for example, $n$ is along the positive $z$-axis, $n^\mu = (1,0,0,1)$, then
one could choose $\bn^\mu =(1,0,0,-1)$ or equally well $\bn^\mu =
(3,2,2,1)$. Since the perpendicular size of the hadron is $\sim 1/\Lambda_{\rm
QCD}$ the momentum $P^\mu$ of a collinear constituent is $(n\cdot P, \bn\cdot
P,P_\perp)\sim Q(\lambda^2,1,\lambda)$, where $\lambda \sim \Lambda_{\rm
QCD}/Q$. This scaling holds for either choice of $\bn$ above. The SCET provides
a systematic way of dealing with the disparate scales $Q\gg \Lambda_{\rm QCD}\gg
(\Lambda_{\rm QCD})^2/Q$. The momentum $P^\mu$ of a fast particle is decomposed
as the sum of a large momentum $p^\mu$ with $\bn\cdot p\sim \lambda^0$,
$p_\perp^\mu\sim\lambda$, and $n\cdot p=0$ and a smaller momentum $k^\mu\sim
\lambda^2$:
\begin{eqnarray} \label{split} 
 P^\mu = p^\mu + k^\mu = \frac{n^\mu}{2}\: \bn\cdot (p+k) + \frac{\bn^\mu}{2}\:
  n\cdot k + (p_\perp^\mu + k_\perp^\mu)\,.  
\end{eqnarray} 
The large momentum $p$ is treated as a label on collinear fields, and the small
residual momentum $k$ is associated with the spatial variation of the fields.
In this paper we show that requiring invariance under the ambiguity in the
decomposition in Eq.~(\ref{split}) has important consequences for collinear
operators in SCET. As examples, we show that this reparameterization invariance
places important restrictions on the form of the leading order collinear quark
action, and fixes the anomalous dimensions of an infinite class of subleading
terms.

The decomposition in Eq.~(\ref{split}) is similar to the one in heavy quark
effective theory (HQET). In HQET $P^\mu = m v^\mu + k^\mu$, where $m$ is
the heavy quark mass, the velocity $v^\mu$ labels HQET fields $h_v(x)$, and
$k^\mu$ is a residual momentum picked out by derivatives on $h_v$. The
ambiguity in the decomposition of $P^\mu$ leads to a reparameterization
invariance~\cite{LM}. This symmetry is the remnant of invariance under the
Lorentz generators $v_\mu M^{\mu\nu}$ which were broken by the introduction of
the vector $v_\mu$ (for the rest frame these generators are the boosts
$M^{0i}=K^i$). Requiring that physics is invariant under the simultaneous change
\begin{eqnarray}
 v_\mu \to v_\mu + \frac{\Delta_\mu}{m}\,,\qquad
 k_\mu \to k_\mu - \Delta_\mu \qquad\quad (v\cdot \Delta=0)
\end{eqnarray}
gives useful constraints on the form of the HQET Lagrangian and
currents~\cite{LM,Chen,FGM,Manohar,Sundrum,Neubert,Campanario:2002fy}.  

In SCET reparameterization invariance is more involved, because the collinear
momentum decomposition in Eq.~(\ref{split}) has a more complicated structure. In
particular, HQET reparameterization invariance only connects operators appearing
at different orders in the $1/m$ expansion, while we will see that the
counterpart in SCET also constrains the form of operators appearing at any given
order.  From Eq.~(\ref{split}) two types of ambiguity are:
\begin{itemize} 
 \item[(a)] The component decompositions, $\bn\cdot(p+k)$ and
$(p_\perp^\mu+k_\perp^\mu)$, are arbitrary by an order $Q\lambda^2$ amount, and
any decomposition should yield an equivalent description.
 \item[(b)] Any choice
of the reference light-cone vectors $n$ and $\bn$ satisfying
\begin{eqnarray} \label{constraints}
 n^2=0 \,,\qquad \bn^2=0 \,,\qquad n\cdot \bn=2 \,,
\end{eqnarray}
are equally good, and can not change physical predictions.
\end{itemize} 
For type (b) the most general infinitesimal change in $n$ and $\bn$ which
preserves Eq.~(\ref{constraints}) is a linear combination of
\begin{eqnarray}\label{repinv}
\mbox{(I)} 
\left\{
\begin{tabular}{l}
$n_\mu \to n_\mu + \Delta_\mu^\perp$ \\
$\bn_\mu \to \bn_\mu$
\end{tabular}
\right.\qquad
\mbox{(II)} 
\left\{
\begin{tabular}{l}
$n_\mu \to n_\mu$ \\
$\bn_\mu \to \bn_\mu + \varepsilon_\mu^\perp$
\end{tabular}
\right.\qquad
\mbox{(III)} 
\left\{
\begin{tabular}{l}
$n_\mu \to (1+\alpha)\, n_\mu$ \\
$\bn_\mu \to (1-\alpha)\, \bn_\mu$
\end{tabular}
\right. \,,
\end{eqnarray}
where $\{\Delta_\mu^\perp,\epsilon_\mu^\perp,\alpha\}$ are five infinitesimal
parameters, and $\bn\cdot\varepsilon^\perp=n\cdot\varepsilon^\perp =
\bn\cdot\Delta^\perp =n\cdot\Delta^\perp= 0$.  Invariance under subset (I) of
these transformations has already been explored in Ref.~\cite{chay}, and used to
derive important constraints on the next-to-leading order collinear Lagrangian
and heavy-to-light currents.  Here we explore the consequences of invariance
under the full set of reparameterization transformations and extend the analysis
of class (I) transformations to higher orders in $\lambda$.  In particular we
show that the transformations in classes (II) and (III) are necessary to rule
out the possibility of additional operators in the lowest order collinear
Lagrangian that are allowed by power counting and gauge invariance.

As might be expected the collinear reparameterization invariance is a
manifestation of the Lorentz symmetry that was broken by introducing the vectors
$n$ and $\bn$. Essentially reparameterization invariance restores Lorentz
invariance to SCET order by order in $\lambda$. The five parameters in
Eq.~(\ref{repinv}) correspond to the five generators of the Lorentz group which
are ``broken'' by introducing the vectors $n$ and $\bn$, namely \{$n_\mu
M^{\mu\nu}$, $\bn_\mu M^{\mu\nu}$\}. If the perpendicular directions are $1,2$
then the five broken generators are $Q_1^\pm = J_{1}\pm K_2$, $Q_2^\pm = J_2\pm
K_1$, and $K_3$.  The type (I) transformations are equivalent to the combined
actions of an infinitesimal boost in the $x$ ($y$) direction and a rotation
around the $y$ ($x$) axis, such that $\bn_\mu$ is left invariant with generators
($Q_1^-$, $Q_2^+$). Type (II) transformations are similar but ($Q_1^+$, $Q_2^-$)
leave $n_\mu$ invariant, while transformation (III) is a boost along the $3$
direction ($K_3$).

In SCET one introduces three classes of fields: collinear, soft and ultrasoft
(usoft), with momentum scaling as $Q(\lambda^2,1,\lambda)$,
$Q(\lambda,\lambda,\lambda)$ and $Q(\lambda^2,\lambda^2,\lambda^2)$,
respectively. For our purposes the interesting fields are those for collinear
quarks ($\xi_{n,p}$), collinear gluons ($A_{n,q}$), and usoft gluons ($A_u$). At
tree level the transition from QCD to collinear quark fields can be achieved by
a field redefinition~\cite{bfps}
\begin{eqnarray} \label{field}
  \psi(x) =  \sum_p e^{-ip\cdot x} \bigg[1 +
  \frac{1}{\bn\cdot {\cal D}}\: \cDslash^\perp\:  \frac{\bnslash}{2}\:
  \bigg]  \xi_{n,p},
\end{eqnarray}
where the two-component collinear quark field $\xi_n$ satisfies~\cite{bfl}
\begin{eqnarray}\label{1}
  \frac{\nslash\bnslash}{4}\: \xi_n =  \xi_n \,, \qquad\quad 
  \nslash\: \xi_n=0 \,.
\end{eqnarray}
The covariant derivatives are further decomposed into two parts, ${\cal D}^\mu =
D_c^\mu + D_u^\mu$, where $D_c^\mu$ and $D_u^\mu$ involve collinear and usoft
momenta and gauge fields respectively. To distinguish the size of derivatives it
is convenient to introduce label operators, $\bnP$ and $\cP_\perp$, which pick
out the labels of collinear fields~\cite{cbis}. For instance, $\bnP \xi_{n,p} =
\bn\cdot p\: \xi_{n,p}$. The label operators allow sums and phases involving
label momenta to be suppressed.  Under gauge transformations the usoft gluons
act like a background field and collinear gauge invariance ensures that only the
linear combinations~\cite{bpssoft}
\begin{eqnarray} \label{cgauge}
 i\bn\cdot D_c = \bnP+g\bn\cdot A_{n,q},\qquad 
 iD_c^\perp=\cP^\perp + g A_{n,q}^\perp,\qquad 
 in\cdot {\cal D} = in\cdot D_u + gn\cdot A_{n,q}
\end{eqnarray}
appear. Each term in Eq.~(\ref{cgauge}) is the same order in $\lambda$. It is
often convenient to swap the order $\lambda^0$ field $\bn\cdot A_{n,q}(x)$ for
the Wilson line
\begin{eqnarray} \label{W}
  W = \Big[ \sum_{\rm perms} \exp\Big( -\frac{g}{\bnP}\: \bn\mcdot A_{n,q}(x) \
  \Big) \Big] = \Big[ \frac{1}{i\bn\cdot D_c}\: \bnP \Big]\,,
\end{eqnarray}
where the differential operators do not act outside the square brackets, and
$[i\bn\cdot D_c\: W]=0$.

Now consider the constraints imposed by reparameterization transformations of
type (a) in Eq.~(\ref{split}). These act only on the components of label
momenta, so
\begin{eqnarray}\label{RIa} 
 \xi_{n,p}(x) \to e^{i\beta\cdot x}\, \xi_{n,p+\beta}(x)\,,\qquad\quad
 A_{n,q}(x) \to e^{i\beta\cdot x}\, A_{n,q+\beta}(x) \,,
\end{eqnarray} 
where the label and phase changes compensate each other. Expressed in terms of
${\cal P}^\mu=\frac12 n^\mu\bnP+\cP_\perp^\mu$, and $i\partial_\mu$ the
transformation in Eq.~(\ref{RIa}) essentially causes
\begin{eqnarray}
  {\cal P}_\mu \to {\cal P}_\mu + \beta_\mu\,,\qquad \quad
  i\partial_\mu \to i\partial_\mu - \beta_\mu\,,  
\end{eqnarray}
with $\beta_\mu=\bn\cdot\beta\, n_\mu/2 + \beta_\mu^\perp$.  This implies that
reparameterization invariant operators must be built out of the linear
combination ${\cal P}_\mu + i\partial_\mu$. Usoft gauge invariance
forces $i\partial_\mu$ to appear along with an usoft gauge field in the
combination $i\partial^\mu + g A_u^\mu$. Furthermore, from Eq.~(\ref{cgauge})
collinear gauge invariance forces collinear fields to appear along with ${\cal
P}_\mu$ and $n\mcdot\partial$. Thus, reparameterization invariant operators
should be built out of
\begin{eqnarray} \label{DD}
 {\cal D}^\mu \equiv D_c^\mu + D_u^\mu \,.
\end{eqnarray}
Since the components of $D_c$ and $D_u$ scale in different ways we see that
Eq.~(\ref{DD}) connects operators at different orders in $\lambda$. It is
important to note, however, that factors of $D_u^\mu$ can also appear by
themselves (when the $\partial_\mu$ acts on non-collinear fields). An example is
the $\Dslash_u$ in the usoft quark Lagrangian.  Eq.~(\ref{DD}) is analogous to
the fact that reparameterization invariance in HQET forces $v^\mu + iD^\mu/m$ to
appear together~\cite{LM}.

The ambiguities of type (b) give more interesting constraints on operators. The
transformations (I,II,III) change the decomposition of any vector or tensor
along the light-like directions. For example, for a vector 
\begin{eqnarray}
  V^\mu =\frac{n^\mu}{2}\, \bn\mcdot V + \frac{\bn^\mu}{2}\, n\mcdot V 
	+ V_\perp^\mu \,,
\end{eqnarray}
the components transform as
\begin{eqnarray}
 \Big(n\mcdot V \,,\, \bn\mcdot V\,,\, V_\perp^\mu \Big) 
  &\stackrel{\rm I}{\longrightarrow}&
  \Big(n\mcdot V + \Delta^\perp\mcdot V^\perp \,,\,
  \bn\mcdot V \,,\, V_\perp^\mu - \frac{\Delta_\perp^\mu}2 \: \bn\mcdot V 
  -\frac{\bn^\mu}2  \Delta^\perp \mcdot V^\perp \Big)\,,\nn\\
 \Big(n\mcdot V\,,\, \bn\mcdot V\,,\, V_\perp^\mu \Big) 
  &\stackrel{\rm II}{\longrightarrow}&
  \Big(n\mcdot V \,,\,\bn\mcdot V+\varepsilon_\perp\mcdot V_\perp \,,\,
  V_\perp^\mu - \frac{\varepsilon_\perp^\mu}2 \: n\mcdot V 
  -\frac{n^\mu}2  \varepsilon_\perp \mcdot V_\perp \Big) \,,\nn\\
\Big(n\mcdot V\,,\, \bn\mcdot V\,,\, V_\perp^\mu \Big) 
  &\stackrel{\rm III}{\longrightarrow}& 
  \Big(n\mcdot V + \alpha n\mcdot V \,,\,\bn\mcdot V-\alpha\bn\mcdot V\,,\,
  V_\perp^\mu \Big)\,.
\end{eqnarray}
These results are easily derived by demanding that the vector itself remains
invariant, $V^\mu\to V^\mu$.  These transformations apply to all vector
components including those of the covariant derivative $\cD^\mu$. Due to the
projection operators in Eq.~(\ref{1}) the transformations also effect collinear
spinors. To derive the spinor transformations we first equate the decomposition
in Eq.~(\ref{field}) before and after the reparameterization transformation,
\begin{eqnarray}
 \sum_p e^{-ip\cdot x} \Big[1 + \frac{1}{\bn\cdot {\cal D}}\: 
 \cDslash^\perp\: \frac{\bnslash}{2} \Big]  \xi_{n,p} = \
 \sum_{p'} e^{-ip'\cdot x} \Big[1 + \frac{1}{\bn\cdot {\cal D}^{\,\prime}}\: 
 \cDslash_{\perp}^{\:\,\prime}\: \frac{\bnslash^\prime}{2} \Big]  \
 \xi_{n,p}^\prime \,,
\end{eqnarray}
where the primes denote quantities transformed under (I), (II), or (III). In each
case multiplying both sides by the transformed $\nslash\bnslash/4$ projection
operator then gives an expression for $\xi_{n,p}^\prime$ in terms of
$\xi_{n,p}$. We find
\begin{eqnarray}
 \xi_n &\stackrel{\rm I}{\longrightarrow}& 
  \left(1 + \frac14\,\Delslash^\perp\bnslash \right)\xi_n \ \,,\\
 \xi_n &\stackrel{\rm II}{\longrightarrow}&
  \left(1 + \frac12\, \epsslash^\perp \frac{1}{\textstyle \bn\mcdot {\cal D}}\: 
  {\cDslash}_\perp \right)\,\xi_n \,,\\
 \xi_n &\stackrel{\rm III}{\longrightarrow}& \xi_n\ \,.
\end{eqnarray}
Finally, we consider the transformation of the Wilson line in Eq.~(\ref{W}).
From this definition we see that $W$ is invariant under type (I) and (III)
transformations. The transformation of $W$ under (II) can be derived by noting
that $[\bn\cdot D_c W]=0$ implies $0=\delta[\bn\cdot D_c W]=[ \delta(\bn\cdot
D_c) W] +[\bn\cdot D_c\: \delta W]$ which we can solve for $\delta W$ to find
\begin{eqnarray}
  W &\stackrel{\rm II}{\longrightarrow}&
  \Big[ \big(1-\frac{1}{\textstyle \bn\mcdot {\cal D}} 
  \,\, \epsilon_\perp\mcdot {\cal D}_\perp \big) W \Big]\,.
\end{eqnarray}
For easy reference a summary of the most commonly used transformations is given
\begin{table}[t!]
\begin{center}
\begin{tabular}{|rcl|rcl|rcl|}
\hline
 \multicolumn{3}{|c|}{Type (I)} & \multicolumn{3}{c|}{Type (II)} 
  & \multicolumn{3}{c|}{Type (III)} \\ \hline
 $n$ & $\to$ & $n+\Delta^\perp$ & $n$ & $\to$ & $n$ & $n$ & $\to$ 
 & $n+\alpha\, n$\\
 $\bn$ & $\to$ & $\bn$ & $\bn$ & $\to$ & $\bn+\varepsilon^\perp$ 
 & $\bn$ & $\to$ &  $\bn-\alpha\, \bn$\\
 \ $n\cdot {\cal D}$ & $\to$ & $n\cdot{\cal D}+\Delta^\perp\mcdot 
 {\cal D}^\perp$ 
  & $n\cdot {\cal D}$ & $\to$ & $n\cdot {\cal D}$ 
  & \ $n\mcdot {\cal D}$ & $\to$ & $n\mcdot {\cal D}+\alpha\, n\mcdot {\cal D}$ 
 \\
 ${\cal D}^\perp_\mu$ & $\to$ & ${\cal D}^\perp_\mu - 
 \mklg{\frac{\Delta^\perp_\mu}2} \: \bn\mcdot {\cal D} - \mklga{\frac{\bn_\mu}2}
 \Delta^\perp \mcdot {\cal D}^\perp$ 
 \ &\  ${\cal D}^\perp_\mu$ & $\to$ & ${\cal D}^\perp_\mu - 
 \mklga{\frac{\varepsilon^\perp_\mu}2}\,  \:
  n\mcdot {\cal D}-\mklga{\frac{n_\mu}2} \, \varepsilon^\perp\! \cdot 
 {\cal D}^\perp$
 \ &\ ${\cal D}^\perp_\mu$  & $\to$ & ${\cal D}^\perp_\mu$  
 \\
 $\bn\cdot {\cal D}$ & $\to$ & $ \bn\cdot {\cal D}$ 
 &\ $\bn\cdot {\cal D}$ & $\to$ & $ \bn\cdot {\cal D} + 
  \varepsilon^\perp\mcdot {\cal D}^\perp$ 
 & \ $\bn\mcdot {\cal D}$ & $\to$ & $\bn\mcdot {\cal D}
 -\alpha\, \bn\mcdot {\cal D}$\:\  
 \\
  $\xi_n$ & $\to$ & $\left(1 + \frac14\,\Delslash^\perp\bnslash \right)\xi_n$&
  $\xi_n$ & $\to$ & $\left(1 + \frac12\, \epsslash^\perp 
  \frac{1}{\textstyle  \bn\mcdot {\cal D}}\: {\cDslash}^\perp \right)\,\xi_n$
 & $\xi_n$ & $\to$ & $\xi_n$
 \\[3pt]
 $W$ & $\to$ & $W$ 
 & $W$ & $\to$ & $\Big[ \big(1-\frac{1}{\textstyle \bn\mcdot {\cal D}} 
  \,\, \epsilon^\perp\mcdot {\cal D}^\perp \big) W \Big]$ 
 & $W$ & $\to$ & $W$\\[4pt]
 \hline
\end{tabular}
\end{center}
{\caption{Summary of infinitesimal type I, II, and III transformations.}
\label{table123} }
\end{table}
in Table~\ref{table123}.

Requiring invariance under the transformations in Table~\ref{table123} places
important constraints on collinear operators. In particular some classes of
operators are ruled out, while others have coefficients whose matching and
anomalous dimensions are related to all orders in perturbation theory.

For the purpose of power counting we take 
\begin{eqnarray} \label{pc}
 \Delta_\perp\sim \lambda\,,\qquad 
 \epsilon_\perp\sim\lambda^{0} \,,\qquad 
 \alpha \sim \lambda^0 \,,
\end{eqnarray} 
and then demand reparameterization invariance order by order in $\lambda$. The
scaling for the infinitesimal parameters is assigned to be the maximum power
that leaves the power counting of the collinear momentum components intact. This
means that only transformations that leave collinear fields collinear are
considered. For example we do not consider large longitudinal boosts $\sim
1/\lambda$ that could turn a collinear momentum into a soft momentum.  For
$\Delta_\perp$ we see from Table~\ref{table123} that the $n$ and $\xi_n$
transformations are suppressed by $\lambda$, while the transformation of $n\cdot
\cD$ and $\cD^\perp$ include homogeneous terms. By homogeneous we mean the same
order in $\lambda$ as the untransformed operator.  For $\cD^\perp_\mu$ the term
$\frac12 \bn^\mu \Delta_\perp\mcdot \cD^\perp\sim \lambda^2$ is smaller than
$\cD^\perp_\mu \sim\lambda$.  For $\varepsilon_\perp$ the transformations of
$\xi_n$, $W$, and $\bn\mcdot \cD$ are suppressed by $\lambda$. The $\cD^\perp$
transformation involves a homogeneous term, as well as a term suppressed by a
single $\lambda$.  Finally, the type (III) transformations are completely
homogeneous.  Note that the homogeneous terms can induce constraints without
mixing orders in the power counting. This occurs for all type (III) cases as
well as for any leading order operator involving a quantity with a homogeneous
term in its type (I) or (II) transformation.

In the remainder of this Letter we present the implications of the symmetry
transformations (I)-(III) for the operators in the collinear quark Lagrangian.
The leading order Lagrange density in the collinear quark sector can be obtained
by tree level matching and is given by \cite{bfps,cbis}
\begin{eqnarray}\label{L0}
{\cal L}_0 = \bar \xi_n \left\{n\cdot iD_u + gn\cdot A_n +
(\SppP + g\Aslash_n^\perp) \frac{1}{\bar{\cal P} + g\bn\cdot A_n}
(\SppP + g\Aslash_n^\perp)\right\} \frac{\bnslash}{2}\: \xi_n \,,
\end{eqnarray}
where ${\cal L}_0\sim \lambda^4$.  As discussed above, invariance under
collinear label reparameterization forces the collinear derivatives in SCET to
appear with the ultrasoft derivatives.  Therefore the Lagrangian (\ref{L0}) is
just the first term in the expansion of the manifestly reparameterization
invariant Lagrangian obtained by replacing the collinear derivatives with
$i{\cal D}$
\begin{eqnarray}\label{L0RI}
{\cal L} &=& \bar \xi_n \left\{n\cdot i{\cal D} +
  i\Dcslash^\perp \frac{1}{\bn\mcdot i{\cal D}}
  i\Dcslash^\perp\right\} \frac{\bnslash}{2} \, \xi_n \,.
\end{eqnarray}
Expanding the derivatives $i{\cal D}$ in powers of $\lambda$, one finds that
${\cal L}= {\cal L}_0 + {\cal L}_1 + {\cal L}_2 + \cdots$, where the terms
${\cal L}_i$ scale like $\lambda^{4+i}$. In particular, the next two subleading
terms in the Lagrange density are
\begin{eqnarray} \label{L12}
{\cal L}_1 &=& \bar \xi_n \left\{
i\Dslash_u^\perp \frac{1}{\bn\cdot iD_c}
i\Dslash^\perp_c  + 
i\Dslash^\perp_c
\frac{1}{\bn\cdot iD_c} i\Dslash_u^\perp\right\}
\frac{\bnslash}{2}\xi_n\\
{\cal L}_2 &=& 
\bar \xi_n \left\{
i\Dslash_u^\perp \frac{1}{\bn\cdot iD_c} i\Dslash_u^\perp -
i\Dslash^\perp_c \frac{1}{\bn\cdot iD_c}
\bn\cdot iD_u \frac{1}{\bn\cdot iD_c} i\Dslash^\perp_c\right\}
\frac{\bnslash}{2}\xi_n \,. \nn
\end{eqnarray}
The expressions in Eq.~(\ref{L12}) agree with the tree level matching result.
The reparameterization argument shows that these terms are connected to the
leading order Lagrangian. Thus, their structure is determined by ${\cal L}_0$ to
all orders in perturbation theory and they have no non-trivial Wilson
coefficients.\footnote{In principle the above results do not rule out
the possibility of additional terms in the power suppressed collinear
Lagrangians which are reparameterization invariant all by themselves.}  This
result for ${\cal L}_1$ agrees with Ref.~\cite{chay}, while the structure of
${\cal L}_2$ (and all higher terms in the expansion of Eq.~(\ref{L0RI})) are
new.
 
The observant reader will have noticed that the reasoning in Eqs.~(\ref{L0})
through (\ref{L12}) relies on the fact that Eq.~(\ref{L0}) is the unique lowest
order Lagrangian.  In fact using only collinear gauge invariance the operator
appearing in Eq.~(\ref{L0}) is not the most general allowed operator. For
instance both
\begin{eqnarray}\label{O1}
 {\cal O}_1 &=& \bar \xi_n\, i\Dslash_c^\perp \frac{1}{\bn\mcdot iD_c} 
   i\Dslash_c^\perp \frac{\bnslash}{2}\, \xi_n\,,\qquad\quad
 {\cal O}_2 =\bar \xi_n\: iD^\perp_{c\mu}\, \frac{1}{\bn\mcdot iD_c}
   \, iD^{\perp\mu}_{c}\, \frac{\bnslash}{2}\, \xi_n \,,
\end{eqnarray}
are collinear gauge invariant, but only ${\cal O}_1$ was included in
Eq.~(\ref{L0}). Even if only ${\cal O}_1$ is present at tree level, the other
operator could in principle be induced through radiative corrections. The
presence of ${\cal O}_2$ at leading order in $\lambda$ would be a concern, since
it would imply that the collinear quark kinetic Lagrangian can only be defined
order by order in perturbation theory. However, it is easy to show that ${\cal
O}_2$ is ruled out by invariance under type (II) reparameterization
transformations in SCET.  We will present this in some detail before extending
the basis in Eq.~(\ref{O1}) to the most general set of gauge invariant operators.

To see that class (I) is not sufficient to rule out ${\cal O}_2$ note that since
$\delta_{\rm (I)}\: [\bnslash\, \xi_n] = 0$ we have
\begin{eqnarray}
 \delta_{\rm (I)}\, {\cal O}_2 &=& -
 \bar \xi_n \: i\Delta^\perp \mcdot D_c\: \frac{\bnslash}{2} \xi_n
 = \delta_{\rm (I)}\, {\cal O}_1\,.
\end{eqnarray}
Here $\delta_{\rm (I)} {\cal O}_{1,2}$ and ${\cal O}_{1,2}$ are the same order in
$\lambda$.  Now at this order we also have
\begin{eqnarray}
 \delta_{\rm (I)}\Big( \bar\xi_n\, n\cdot i {\cal D}\:\frac{\bnslash}{2} 
  \xi_n\Big)
   = \bar \xi_n \: i\Delta^\perp \mcdot {D_c}\: \frac{\bnslash}{2} \xi_n\,.
\end{eqnarray}
This shows that reparameterization invariance of type (I) ties the first two
terms in Eq.~(\ref{L0}) with the third one.  However, as far as transformations
of type (I) are concerned we could replace ${\cal O}_1$ in Eq.~(\ref{L0}) by
${\cal O}_2$ and still leave ${\cal L}_0$ invariant.

For a type (II) transformation we begin by noting that for any scalar operator
in the single collinear quark sector the terms homogeneous in $\lambda$ vanish
identically. This follows from the fact that the index on $D_\perp^\mu$ is
contracted into another perpendicular vector, and that the transformation for
$\bn^\mu$ is subleading except for when it is contracted with $\gamma_\mu$. As
discussed in Ref.~\cite{bfprs} a complete Dirac basis for these bilinears is
$\bnslash$, $\bnslash\gamma_5$, and $\bnslash\gamma_\perp^\mu$. However, if
$\bnslash$ in any of these Dirac structures is transformed the resulting
structure vanishes between $\xi_n$ fields. This implies that as far as the
collinear Lagrangian is concerned we can impose invariance under the remaining
type (II) transformations order by order in $\lambda$ without worrying about
connecting terms of different orders. In general this need not be true for
currents.

Making a type (II) transformation one finds that the change in the first term of
the effective Lagrangian in Eq.~(\ref{L0}) is
\begin{eqnarray}
& &\delta_{\rm (II)}\ \bar\xi_n n\cdot i{\cal D}  \frac{\bnslash}{2}\xi_n = 
 \bar \xi_n \left( i\Dslash_c^\perp \frac{1}{\bn\mcdot i D_c} 
 \frac{\epsslash^\perp}{2} n\cdot i{\cal D} +
 n\cdot i{\cal D} \frac{\epsslash^\perp}{2} \frac{1}{\bn\mcdot iD_c} 
 i\Dslash_c^\perp \right)\frac{\bnslash}{2}\xi_n \,.
\end{eqnarray}
This variation is exactly canceled by the change in ${\cal O}_1$ coming from the
$i\Dslash_c^\perp$ factors. The remaining variation in ${\cal O}_1$ is from
the change in the $\xi_n$ fields and in the $1/\bn\cdot iD_c$ factor, thus
\begin{eqnarray} \label{varyII1}
 \delta_{\rm (II)} {\cal O}_1 &=& 
 -\delta_{\rm (II)}\ \bar\xi_n n\cdot i{\cal D} \,\frac{\bnslash}{2}\xi_n 
 + \bar \xi_n \left\{ i\Dslash_c^\perp \frac{1}{\bn\cdot iD_c} 
 \frac{\epsslash^\perp}{2} i\Dslash_c^\perp 
 \frac{1}{\bn\cdot iD_c} i\Dslash_c^\perp\right.\\
& &\quad\left. - i\Dslash_c^\perp \frac{1}{\bn\cdot i D_c} 
  \: \varepsilon_\perp\mcdot i{D}_c^\perp \frac{1}{\bn\cdot i D_c} 
  i\Dslash_c^\perp + i\Dslash_c^\perp \frac{1}{\bn\cdot iD_c} i\Dslash_c^\perp
  \frac{\epsslash^\perp}{2} \frac{1}{\bn\cdot iD_c} i\Dslash_c^\perp\right\}
  \frac{\bnslash}{2}\xi_n \nn\\
&=& -\delta_{\rm (II)}\ \bar\xi_n n\cdot i{\cal D} \,\frac{\bnslash}{2}\xi_n 
 \,.\nn
\end{eqnarray}
On the other hand, the variation of the $D_\perp$'s in ${\cal O}_2$ gives terms
with $\epsilon^\perp\mcdot D_c^\perp$ which can not cancel the variation of the
$in\mcdot {\cal D}$ term. Furthermore, for ${\cal O}_2$ the terms analogous to
the ones in curly brackets in Eq.~(\ref{varyII1}) do not vanish because the
first and third terms from the variation of $\xi$ do not cancel against the term
from the $i\bn\mcdot D$ variation.  Thus, ${\cal O}_2$ is not allowed in the
leading order effective Lagrangian by reparameterization invariance of type
(II).

We now proceed to a more general analysis of the complete set of gauge invariant
operators in the collinear Lagrangian up to $\lambda^4$. A special feature of
SCET is the presence of the dimensionful collinear covariant derivative
$\bn\cdot iD_c = \bar {\cal P} + g\bn\cdot A_n$ which scales like
$\lambda^0$. Because of this operator, dimensional analysis and gauge invariance
are not sufficient to completely constrain the operators that appear at a given
order in $\lambda$.  Since we require two $\xi_n$ fields the operators in the
Lagrange density start at order $\lambda^2$. At this order gauge invariance
allows
\begin{eqnarray} \label{Lm2}
  {\cal L}_{-2} &=& \bar \xi_n\: {\bn\mcdot i{\cal D}}\, \frac{\bnslash}{2}\: 
  \xi_n\,.
\end{eqnarray}
However, invariance under transformation (III) requires that each time $\bn$
appears in the numerator it is accompanied by either an $n$ in the numerator or
a factor of $\bn$ in the denominator. Thus, the Lagrangian in Eq.~(\ref{Lm2}) is
not allowed by type (III) reparameterization invariance. At order $\lambda^3$
the only way to construct an operator is with one factor of $D_\perp^\mu$.
Since the Lagrangian is a scalar this index must be contracted with an object
that will not increase the power of $\lambda$, which leaves $\bar\xi_n \bnslash
\cDslash_\perp \xi_n$. However, this operator is also ruled out by type III
reparameterization invariance.

At order $\lambda^4$ there are infinitely many operators allowed by gauge
invariance and invariance under (III):
\begin{eqnarray}\label{L0gen}
 {\cal L}^\prime_0 &=&  \bar\xi_n\: {\cal O}\: \frac{\bnslash}{2}\: \xi_n \, , 
\end{eqnarray}
where 
\begin{eqnarray}
 {\cal O} &=& \sum_a c_a {\cal O}^a + \sum_{a,b}(s_{ab} {\cal S}^{ab} +
  t_{ab} {\cal T}^{ab})\,,
\end{eqnarray}
and the operators ${\cal O}^a, {\cal S}^{ab}$ and ${\cal T}^{ab}$ are
\begin{eqnarray} \label{genL}
{\cal O}^a &=&  N^a (n\mcdot i{\cal D}) \frac{1}{N^a} 
  + \frac{1}{N^a}(n\mcdot i{\cal D}) N^a \, ,\\
{\cal S}^{ab} &=&  N^a i\cDslash_\perp \frac{1}{N^{a+b+1}} 
  i\cDslash_\perp N^b + N^b i\cDslash_\perp \frac{1}{N^{a+b+1}} 
  i\cDslash_\perp N^a \, , \nn \\
{\cal T}^{ab} &=& N^a i{\cal D}_{\perp}^\mu \frac{1}{N^{a+b+1}} 
 i{\cal D}_{\mu}^{\perp}N^b  + N^b i{\cal D}_{\perp}^\mu \frac{1}{N^{a+b+1}} 
 i{\cal D}_{\mu}^\perp N^a \, . \nn
\end{eqnarray}
To simplify the formulae we have defined $N \equiv \bn\cdot i{\cal D}$.  The
operators ${\cal O}^a, {\cal S}^{ab}$ and ${\cal T}^{ab}$ are Hermitian so the
coefficients $c_a, s_{ab}$ and $t_{ab}$ are real numbers. Since $[iD_\perp^i,
iD_\perp^j] = i g\, G_\perp^{ij}$ and $[N,iD_\perp^i]= i g\, \bn_\mu G^{\mu i}$,
any operator that is order $\lambda^4$ and contains a collinear gluon field
strength can be reduced to a linear combination of those in Eq.~(\ref{genL}).

The variation of Eq.~(\ref{L0gen}) under the transformation (II) is given by
\begin{eqnarray}\label{dL}
  \delta_{\rm (II)}{\cal L}^\prime_0 
 = \bar\xi_n \left[\delta_{\rm (II)}{\cal O} 
  + i\cDslash_\perp \epsslash^\perp \frac{1}{2 N}{\cal O} 
  + {\cal{O}}\frac{1}{2 N}\epsslash^\perp i\cDslash_\perp \right] 
  \frac{\bnslash}{2}\xi_n\, .
\end{eqnarray}
Most terms appearing in Eq.~(\ref{L0gen}) will not be invariant unless the
operator coefficient is set to zero. First consider the variation of ${\cal
S}^{ab}$:
\begin{eqnarray}\label{dS}
 \delta_{\rm (II)} {\cal S}^{ab} &=& 
 [\delta_{\rm (II)} N^a] i\cDslash_\perp \frac{1}{N^{a+b+1}} 
 i\cDslash_\perp N^b + N^a i\cDslash_\perp \frac{1}{N^{a+b+1}} i\cDslash_\perp 
 [\delta_{\rm (II)} N^b] \\
&&- \frac{1}{2} N^a \epsslash_\perp n \mcdot i {\cal D} \frac{1}{N^{a+b+1}} 
 i \cDslash_\perp N^b - \frac{1}{2} N^a \cDslash_\perp \frac{1}{N^{a+b+1}} 
 i \epsslash^\perp n\mcdot i {\cal D} N^b \nn \\
&& + N^a i \cDslash_\perp \left[\delta_{\rm (II)}\frac{1}{N^{a+b+1}} \right] 
 i\cDslash_\perp N^b + (a \leftrightarrow b) \,,  \nn
\end{eqnarray}
where for positive $a$
\begin{eqnarray}
\delta_{\rm (II)}N^a &=& \epsilon^\perp \cdot i {\cal D}^\perp N^{a-1} 
 + N \epsilon^\perp \cdot i {\cal D}^\perp N^{a-2} + ... 
 + N^{a-1} \epsilon^\perp \cdot
i {\cal D}^\perp 
\, , \\
\delta_{\rm (II)}\frac{1}{N^a} &=& -\frac{1}{N}\,\epsilon^\perp \mcdot i D^\perp 
 \frac{1}{N^{a}} -\frac{1}{N^2}\, \epsilon^\perp \mcdot i {\cal D}^\perp 
 \frac{1}{N^{a-1}} - ...-  \frac{1}{N^a}\, \epsilon^\perp \mcdot 
 i {\cal D}^\perp \frac{1}{N} \, . \nn
\end{eqnarray}
The vanishing of Eq.~(\ref{dL}) requires $\delta_{\rm (II)}{\cal O}$ to consist
of operators in which $\cDslash^\perp$ appears all the way to the left or all the
way to the right.  Except for the case $a=b=0$, the operators appearing in
$\delta_{\rm (II)}{\cal S}^{ab}$ are not of this form. These operators cannot be
canceled by variations of ${\cal O}^a$ or ${\cal T}^{ab}$ since the latter
have a trivial Dirac structure, while the operators in $\delta_{\rm (II)}{\cal
S}^{ab}$ do not. Thus invariance under (II) requires $s_{ab}=0$ except for
$s_{00}$, which will be set to $1/2$ once the normalization of the free quark
kinetic term is fixed.

The above analysis implies that  
\begin{eqnarray}
  {\cal O} = i\cDslash^\perp\frac{1}{N} i\cDslash^\perp +{\cal O}^\prime \, ,
\end{eqnarray}
where ${\cal O}^\prime$ contains contributions to ${\cal O}$ from the operators
${\cal O}^a$ and ${\cal T}^{ab}$. Inserting this into Eq.~(\ref{dL}), one finds
after some algebra:
\begin{eqnarray}
 \delta_{\rm (II)}{\cal L}^\prime_0 
 = \bar\xi_n \left[\delta_{\rm (II)}\ {\cal O}^\prime 
  + i\cDslash^\perp \epsslash^\perp \frac{1}{2 N}({\cal O}^\prime 
  - n\mcdot i {\cal D}) + ({\cal O}^\prime - n\mcdot i {\cal D})\frac{1}{2 N}
  \epsslash^\perp i\cDslash^\perp \right] \frac{\bnslash}{2}\xi_n\, .
\end{eqnarray}
The terms involving $\delta_{\rm (II)}{\cal O}^\prime$ and $\epsslash^\perp$
must vanish independently since they have different Dirac structures. Thus,
${\cal O}^\prime = n\cdot i D$ and $\delta_{\rm (II)}{\cal O}^\prime = 0$ which
together imply $\delta_{\rm (II)}{\cal L}^\prime_0=0$.  Thus, reparameterization
invariance of type (II) and (III) require the Lagrangian to be ${\cal L}_0$ to
all orders in perturbation theory. Since the free quark kinetic terms come from
${\cal L}_0$ the normalization of these terms is entirely fixed and they cannot
acquire an anomalous dimension. This completes the proof that the result in
Eq.~(\ref{L0RI}) is the most general reparameterization invariant collinear
quark Lagrangian whose expansion starts out at order $\lambda^4$.

In this paper we have used reparametrization invariance to constrain the form of
the collinear SCET Lagrangian. An important result of this paper is that the
form of the leading Lagrangian is uniquely determined by gauge invariance,
reparameterization invariance, and tree level matching.  Reparametrization
invariance was also used to identify the complete set of subleading operators
containing usoft gluon fields which are related to the leading order Lagrangian,
and therefore are not renormalized to all orders in perturbation theory. We have
not been able to construct subleading corrections to the two quark sector of the
collinear Lagrangian which are by themselves reparameterization invariant. An
important open question is whether such operators exist. If not, then
reparameterization invariance would uniquely determine the SCET Lagrangian in the
two collinear quark sector.  It is of course possible to write down operators
coupling to external currents which involve one or more collinear quark fields
and receive nontrivial renormalization. Examples include the heavy to light
currents~\cite{bfl,bfps,chay} and deep inelastic scattering~\cite{bfprs}. The
anomalous dimensions of these operators can be used to sum logarithms involving
ratios of the hard and infrared scales.

An application of reparameterization invariance which we have not considered in
this paper is matching and anomalous dimension calculations for operators. For
processes with energetic hadrons a local QCD current matches onto an infinite
series of operators in SCET. Reparameterization invariance can be used to derive
relations between Wilson coefficients of different operators in this expansion
that are valid to all orders in perturbation theory much like the relations for
heavy-to-heavy currents in HQET~\cite{Neubert}. In SCET relations have been
derived for heavy-to-light currents in Ref.~\cite{chay} using the type (I)
transformations. It would also be interesting to extend these calculations to
include type (II) transformations as well as currents composed entirely of
collinear fields. The latter should have applications for the calculation of
power suppressed cross sections in high energy processes.

\begin{acknowledgments}
This work was supported in part by the Department of Energy under the grants
DOE-FG03-97ER40546 and DE-FG03-00-ER-41132. T.M. is supported by DOE grant
DE-FG02-96ER40945 and by the Thomas Jefferson National Accelerator Facility,
which is operated by the Southeastern Universities Research Association (SURA)
under DOE contract DE-AC05-84ER40150. T.M. also thanks the UCSD theory group
for their hospitality.

\end{acknowledgments}


\end{document}